\documentclass[10pt, conference, compsocconf]{IEEEtran}
\ifCLASSINFOpdf
\else
\fi
\usepackage{amsmath}
\usepackage{graphicx}
\usepackage{caption}
\usepackage{refstyle}
\usepackage{color}
\usepackage{flushend}
\usepackage{balance}
\usepackage{url}

\begin{document}

	%\title{Find Optimal Checkpoint Interval for Inter-dependent parallel processes running on Volunteer PC grids}
		
	\title{Checkpointing to minimize completion time for Inter-dependent Parallel Processes on Volunteer Grids}

	\author{
		\IEEEauthorblockN{Mohammad Tanvir Rahman \,\,\,\, Hien Nguyen \,\,\,\, Jaspal Subhlok \,\,\,\, Gopal Pandurangan}
		\IEEEauthorblockA{Department of Computer Science, University of Houston \\
			Houston, TX 77204, USA \\
			Email: mtrahman3@uh.edu, hxnguyen4@uh.edu, jaspal@uh.edu, gopal@cs.uh.edu}
	}
	
	\maketitle

	\begin{abstract}
	
Volunteer computing is being used successfully for large scale scientific computations. This research is in the context of Volpex, a programming framework that supports communicating parallel processes in a volunteer environment. Redundancy and checkpointing are combined to ensure consistent forward progress with Volpex in this unique execution environment characterized by heterogeneous failure prone nodes and interdependent replicated processes. An important parameter for optimizing performance with Volpex is the frequency of checkpointing.
The paper presents a mathematical model to minimize the completion time for inter-dependent parallel processes running in a volunteer environment by finding a suitable checkpoint interval. Validation is performed with a sample real world application running on a pool of distributed volunteer nodes. The results indicate that the performance with our predicted checkpoint interval is fairly close to the best performance obtained empirically by varying the checkpoint interval. 

%While a volunteer computing environment is the context of this research, the results are applicable to broader distributed computing where the goal is to achieve good application performance in the presence of frequent failures.

\end{abstract}


	\IEEEpeerreviewmaketitle
	\section{Introduction}
	%\textbf{1.	What is the problem?\\}
	The goal of the research presented in this paper is to minimize the completion time for inter-dependent parallel processes running in a volunteer environment. The pool of hosts in a volunteer environment consists of distributed computers which can be used for performing large scale computations when ``idle''. Here ``idle'' implies the conditions under which the owner of the machine allows his/her computer's use for volunteer computing. The computers can be geographically distributed and connected through the internet.
	At present,  volunteer computing is being used successfully for large scale scientific computations, commonly employing BOINC~\cite{anderson04}. Current large volunteer communities can provide computation power in the range of Petaflops. This CPU power is completely free and comes from volunteer idle computers located all over the globe. The parallel processes running on different volunteer clients can coordinate and communicate with each other with Volpex~\cite{Hien} which provides a custom-defined  put/get model for communicating processes.  
	
	%The state of 	``idle" is user-defined. A computer (also known as client) can be marked as idle if its CPU usage goes below a certain threshold. Each idle client can run one (in case of single-core machine) or more (for multi-core machine) parallel processes. The parallel processes running on different clients can exchange data for the sake of computation. 
	
	One of the key characteristics of volunteer computing is that the hosts are highly unreliable as they can leave the available pool simply because the owner turns them off or starts using them. Periodic checkpointing and maintaining redundant replicas are important for making continuous forward progress in a volunteer environment. In this paper, our goal is to minimize the overall completion time by predicting a suitable checkpoint interval in the presence of process replication. 
	%The checkpoint interval is an important performance parameter for volunteer computing and the state-of-the-art does not provide any solution to the optimization of replicated parallel processes. 
	Hence, this research is important for effective use of volunteer nodes for large parallel applications. Our  proposed solution is deployed and evaluated on a pool of volunteer nodes managed by BOINC and Volpex.
	%However, the approach can be extended to any distributed environment with multiple processes prone to failures. 
		
	%\textbf{2.	Why is it interesting and important?\\}
	
	%The more we minimize the completion time, the more throughput we gain from the system. Since our proposed algorithm can be extended to a distributed network, this may also contribute to maximizing system throughput of a distributed system.

	%\textbf{3.	Why is it hard? (E.g., why do naive approaches fail?)\\}
	Creating checkpoints is an effective technique to deal with failures. It is important for minimizing  completion time and increases collective throughput. However, creating checkpoints consumes time and resources. Making checkpoints too frequently will waste a lot of time and resources, but on the other hand, making infrequent checkpoints will result in a significant amount of work-loss in case of client failures. The presence of redundant replicas of individual processes makes the optimal checkpoint interval calculation more complex. To the best of our knowledge, state-of-the-art in checkpoint optimization does not provide any solution for replicated coordinated parallel processes.

	%\textbf{4.	Why hasn't it been solved before? (Or, what's wrong with previous proposed solutions? How does mine differ?)\\}
	There is a large body of research on optimizing the checkpoint interval, some examples being~\cite{Hursey:AgosticCPIOpenMPI,Priya:CP,yudan:OptCPrestartModel}. 
	%However, finding optimal checkpoint interval for  inter-dependent parallel processes with replication is not well-explored yet.
	However, most of the earlier work focuses on finding the optimal checkpoint for a single process or independent distributed processes. Presence of communicating processes fundamentally changes the execution model as the slowdown or failure of a single process impacts the entire computation. Presence of replicated processes further complicates the execution model, as the computation can continue seamlessly despite failures as long as at least one replica of each process is alive.
	%Our Volpex ~\cite{Hien} framework splits high computation tasks into multiple chunks and then uses a custom-defined shared put/get method for communicating among parallel processes.  
	Current Volpex framework uses a heuristic predefined checkpoint interval for application execution that is based on the programmer's intuition. Work presented in this paper automates the process of checkpoint interval selection for  multiple inter-dependent parallel processes, each possibly with multiple replicas. 
	%That's why this approach is believed to be better and more efficient.

	%\textbf{5.	What are the key components of my approach and results? Also include any specific limitations.\\}
	
	The mathematical model developed for computing the optimal checkpoint interval uses four input parameters: the number of processes, the number of replicas for each process, time to create a checkpoint, and the process success distribution (distribution that gives the probability that the process will survive till a given time). The model allows the calculation of the theoretically optimal checkpoint interval. The checkpoint prediction model was validated in Volpex by running  a real world application with varying number of processes, levels of replication and checkpoint size. The results show that the predicted checkpoint interval calculated by our analysis  yields  a job completion time that is close to the minimum.  
	%does not always provide the lowest completion time because of the assumptions in the system, but effectively minimizes the job completion time in practice.

	The proposed approach requires success distribution of computation hosts as input. To find an accurate success distribution, we analyzed client availability over a period of time for the pool of volunteer hosts available to us. This approach is time consuming which is a limitation of our approach. On the other hand, prior knowledge regarding the process failures in similar settings can be used for estimating the optimal checkpoint interval.
	
	To summarize, our main contributions are:
	
	\begin{enumerate}
		\item Designing a probability based mathematical model to predict a checkpoint interval with the lowest completion time for inter-dependent parallel processes with multiple replicas. The true optimal checkpoint interval is unknown and can only be estimated empirically.
		\item	Implementation  and evaluation of the checkpoint prediction model in a real world volunteer computing framework with a real world application to validate  
		our theoretical analysis.
	\end{enumerate}

	\section{Previous Work}

	\textbf{Simple optimum checkpoint interval for a single process: } Young~\cite{Young:1974_FirstOrder} initiated the work on approximating the optimum checkpoint interval for minimizing application run time. He gave a first order approximation to the optimum checkpoint interval for a single process as:
	\begin{equation*}
		T_c^{opt} \approx \sqrt{2T_s*T_f}
	\end{equation*}
	where $T^{opt}_c$= Time interval between Checkpoints, $T_s$= Time to save information at a Checkpoint, and $T_f$ = Mean Time to  Failure (MTTF). This is a simple mathematical equation to find the optimal checkpoint interval and very easy to apply in practice. However, Young did not consider multiple parallel processes in his work. Our approach  is very similar to Young's approach (for one process), but we do a simpler analysis, since we assume that when a process fails it has to wait till checkpointing to restart. 
	In fact, when we use a crude (first-order) approximation, we get  $T_c^{opt} \approx \sqrt{T_s *T_f}$  (see Section IV.A) which is essentially the same as Young's approximation (but without the factor of $\sqrt{2}$).
	The advantage of our approach is that it is simpler, and also leads to more accurate equation for $T_c^{opt}$ (under the assumptions) and generalizes easily for multiple processes and many replicas.
	
	\textbf{Higher order optimum checkpoint interval for a single process: } Daly discussed three models for predicting the runtime and optimum restart	interval~\cite{Daly:2003ModelForPredicting}. He evaluated these models and used the results to	derive a simple method for calculating the optimum restart interval. He showed that 
	$T_c^{opt} = \sqrt{2 \delta (M+R)} - \delta$  is an excellent estimator of the optimum compute interval between restart dumps for values of $(T_c^{opt}+\delta)/M < \frac{1}{2}$,
	where, $M$ = mean time between system failures, $\delta$ =  time to write a checkpoint and $R$ = Restart overhead.
	Daly continued the work further and  developed a higher order model for finding the optimum restart interval on a system exhibiting Poisson single component failures ~\cite{Daly:Higher_Order_Estimate}. He derived a complete cost function and demonstrated a perturbation solution that provides accurate high order approximations (99.8\% of the exact solution time) to the optimum checkpoint interval. 
	%He demonstrated solution that provided accurate high order approximations (99.8\% of the exact solution time) to the optimum checkpoint interval. 
	However, checkpoint interval for inter-dependent parallel processes are not discussed in Daly's work.
	
	\textbf{Multi-level checkpointing (SCR): } Moody et al. ~\cite{Moody:2010:DME:1884643.1884666} designed Scalable Checkpoint/Restart(SCR), a multi-level checkpoint system that writes checkpoints to RAM, Flash or disk on the compute node in addition to the parallel file system. They developed low-cost checkpointing schemes that are 100x-1000x faster than a parallel file system for MPI processes. However, the problem we are trying to solve is in the context of Volpex Dataspace API which is a Put/Get communication model. MPI processes with different level of storage hierarchy simply do not apply in this framework. Moreover, our problem considers process replication which is not considered in this work. 
	
	\textbf{Impact of parameters:} Zheng et al.~\cite{subhlok_Zheng} summarized failure and availability patterns of volatile distributed computing systems. The authors proposed simple models for characterizing the impact of parameters on the efficiency of checkpointing with restart and replication schemes. They showed that when the number of processors and/or the failure rate is high, it is beneficial to use a replication scheme rather than a checkpoint-restart scheme. However, there is no formulation for calculating optimal checkpoint interval for inter-dependent parallel processes in their work.
		
	\textbf{Host Availability Analysis:} One of the key inputs for measuring the optimal checkpoint in a volunteer environment is ``success distribution" of the system. Success distribution is dependent on system's host availability. Several projects have addressed  finding host availability of a system~\cite{Kondo:correlated,EsTA09:PerfPrediction,Andr:Exploiting}. Although there is no straightforward way of determining host availability in a heterogeneous environment, profile modeling can give a rough estimate of host availability in a volunteer system. In profile modeling, the client availability pattern~\cite{DanielKondo:Longterm} is studied to understand when a computer will be available for performing a volunteer job. Kondo et al.~\cite{Kondo:Seti@home} described an effective method for classifying subsets of hosts' availability in a large distributed heterogeneous system. These methods and models are critical for the design of stochastic scheduling algorithms across large systems where host availability is uncertain. The process of finding host availability for a large heterogeneous system is out of the scope for this paper. We assume that the success distribution will be provided as an input parameter at the beginning of program execution. 
	%To estimate the success distribution of the system we assumed that all the nodes in the system have same availability distribution.
	Moreover, using workload data and simulation, Jones  et al.~\cite{Jones:2012AppMonitoring} showed that application efficiency is fairly insensitive to error in estimation of application's mean time to interrupt (AMTTI) or, in other words, mean time to failure.

\section{Execution Model}
\label{sec:model}

The research in this paper is motivated by the challenges faced in application execution under Volpex. We provide a brief introduction to the Volpex execution framework as that also forms the context for the results developed in this paper.

Volpex employs BOINC~\cite{Anderson:2006:Designing_Runtime_System,anderson04}, a well-known middleware for running scientific applications over volunteer nodes, for the basic control and management of execution. Volpex~\cite{Hien} provides a programming framework and runtime support for the execution of communicating parallel codes on volunteer nodes. Volpex has developed two programming frameworks: the Dataspace API~\cite{kanna:2010communication, Rohit:RobustComm} with a Put/Get communication model and an implementation of a subset of the MPI standard called the VolpexMPI~\cite{anand10:CommTargetSelection, Anand:2011:RobustEfficientmsgPassing}. Volpex uses BOINC middleware to recruit and manage volunteer nodes to run the application. The number of hosts to be recruited depends on the number of processes, level of replication, the number of requested spare hosts, and other user-defined parameters. During the execution of the application, new process instances can be created by selecting nodes from a pool of ``hot spares''. Additional nodes can be recruited during execution when the pool of spares is diminished to a preset value because of host failures.

In the Volpex framework, an application executes with $n$ processes, each with $r$ concurrent executing replicas. Volpex can explicitly create process replicas during program invocation or a new instance may be created from a checkpoint to replace a slow or failed process instance. Process replicas are not aware of the existence of, or coordinate with, other process replicas and usually act autonomously. A process can checkpoint its state independently without the coordination of other processes. The system can recreate a process from a checkpoint irrespective of whether the original process is dead or alive, and without coordinating with other replicas or processes. Volpex uses ``heartbeat monitoring" to detect process failure.

In a checkpointing Volpex application, each process instance makes a \textit{StoreCheckpoint} request to the server after performing a fixed amount of work (for example, finishing a fixed number of loop iterations), rather than after a fixed time period. A fast client will send this request more frequently than a slower machine. The server determines if the \textit{StoreCheckpoint} request will lead to the recording of an actual checkpoint. A \textit{StoreCheckpoint} request normally leads to an actual recording of a checkpoint, if: (1) it represents the most recent checkpoint state, i.e., there does not already exist a checkpoint on the server that represents a process state later in logical time (this is possible with replicas) and (2) a specified {\em checkpoint interval} has elapsed since the recording of the last checkpoint. If these conditions are satisfied, the server will accept that \textit{StoreCheckpoint} request and store a checkpoint on the server. Otherwise, the server will ignore the request. The pseudo-code for checking conditions to store a checkpoint in Volpex is as follows:	
	
	\begin{table} [h]
		\centering
		\begin{tabular}{|l|}
			\hline
				\noindent If (requested checkpoint number $>$ last saved checkpoint number and \\ 
				elapsed time from last saved checkpoint $>$ checkpoint interval)\\
				\hspace{6 mm} server saves the checkpoint\\
				Else\\
				\hspace{6 mm} server ignores the \textit{StoreCheckpoint} request\\
			\hline
		\end{tabular}
		\label{tab:execution_with_checkpoint_proceeds}
	\end{table}	
	
Each application is configured to execute with a fixed number of replicas for each process. When all replicas of a process are dead (less common), a new process is created immediately from the most recent checkpoint on the server. When a process replica is dead but other replicas remain (more common), a new instance of the process is created after getting the next valid \textit{StoreCheckpoint} request from another replica. The pseudo-code for checking conditions to start a process after client failure in Volpex is as follows:

	\begin{table} [h]
		\centering
		\begin{tabular}{|l|}
			\hline
			If there is no other replica\\
			\hspace{6 mm} server immediately creates a thread starting from latest checkpoint\\
			If there is another alive replica,\\ 
			\hspace{6 mm} server waits till the next \textit{StoreCheckpoint} request, \\
			\hspace{6 mm} then creates a new replica thread starting from latest checkpoint\\
			\hline
		\end{tabular}
		\label{tab:client_failure}
	\end{table}
	
By default, Volpex applications have to be configured with a heuristic fixed {\em checkpoint interval} for all the processes as input before the execution begins. The goal of this work is to determine the checkpoint interval automatically for applications that optimizes performance.

%Gopal --- Use the terminology "checkpoint" throughout.

	\section{A mathematical model and analysis}

In this section, we develop the mathematical framework for predicting the optimal checkpointing interval. We assume a simple probabilistic model for process failures and analyze the optimal checkpoint interval under this assumption. Following  are the input parameters:

	\begin{enumerate}
		\item Number of processes: $n$
		\item Number of replicas for each process: $r$
		\item Time to create a checkpoint: $T_s$. $T_s$ depends on various factors like checkpoint size, network congestion, link bandwidth, client/server configuration etc. We used the following cost function to calculate the $T_s$.
		\begin{equation}
			T_s = T_{s(cl)} + T_{s(lat)}+T_{s(up)} +T_{s(ack)} \label{Ts formula}
		\end{equation}
		where, 
		$T_{s(cl)}$ = time to create checkpoint in client, $T_{s(lat)}$ = time to send checkpoint from client to server, $T_{s(up)}$ = time to upload checkpoint to server and $T_{s(ack)}$= time to send acknowledgment to client.
		\item Success probability distribution (of a single process with no replica): 	
			$p = f(t)$.	
		The success distribution gives, for a single process having no replica, the probability $p$ (which depends on $t$) to reach a checkpoint without failure.  In other words, the probability of a process being alive till time $t$ is $f(t)$.  
		In this work, we will  assume the success probability distribution to be the {\em exponential distribution} function \cite{Ross}.
		The exponential distribution is commonly used to model failures in similar settings (e.g.,  \cite{Ross,Young:1974_FirstOrder}). In particular, it is also	used by Young \cite{Young:1974_FirstOrder} to model process failures. The main property of the exponential
		distribution is the {\em memorylessness} property: given that there is no failure till time $t$, the probability of failure till time $t+s$ is exactly the probability of failure till time $s$, i.e., $f(t+s)$, given that there is no failure of the process till time $t$,
		is $f(s)$. An equivalent way is to assume  that the occurrence of failures is a  {\em Poisson}  process with failure rate $\lambda$. In other words, failures are assumed to occur independently and at random.
		 The success probability distribution of the exponential distribution is defined as
	\begin{equation}
	 \label{cdf-expo} 
		f(t) = e^{-\lambda t}	
		\end{equation}
	where $\lambda$ is the failure rate or $\frac{1}{\lambda}$ is the mean time to failure (MTTF).
	\end{enumerate}
	
	Our goal is to compute $T^{opt}_c$, the optimum checkpoint interval. To do this we will compute the expected time incurred
	(under the given success distribution) to successfully reach a checkpoint (including the time to create a checkpoint).
	Let's denote this expected time by $G(T_c)$, which is a function of $T_c$. $G(T_c)$ can be considered  the (expected) overhead  associated with having checkpoint interval as $T_c$ under success distribution $p$.   The ``normalized" overhead is $G(T_c)/T_c$, i.e., the ratio of the expected overhead time to the actual checkpoint interval ($T_c$ is the time when ``useful" work is done). This captures the amount of overhead incurred
	per  checkpoint interval. 
	The optimum checkpoint interval $T^{opt}_c$ is the checkpoint interval
	that minimizes $G(T_c)/T_c$.

	%Given the success probability distribution, 
	
	\iffalse
	\textbf{Output:} Let $T_c$ be the checkpoint interval. Average time to reach a checkpoint 
	is the Expected time  to reach a checkpoint, denoted as $E(T_c)$
	\begin{equation*}
		\text{Then, }overhead(T_c) = \frac{E(T_c)}{T_c}
	\end{equation*}
	As an output, We need to find the optimal checkpoint time which has minimum overhead. That is
	\begin{equation}
		T_c(op) = min[overhead (T_c)]; \text{where } T_c = 1...\infty
	\end{equation} 
	\fi
	
	\iffalse
	\begin{table} [h]
			\centering
			\caption{Notations}
			\begin{tabular}{|l|l|}
				\hline
				\textbf{Notation} & \textbf{Meaning} \\
				\hline
				$T_c$ & Checkpoint interval \\
				\hline
				$T_s$ & Time to create Checkpoint \\
				%\hline
				%$T_r$ & Time to detect fault and restart \\
				\hline
				$n$ & Number of processes \\
				\hline
				$r$ & Number of replica for each (single) process \\
				%\hline
				%$E(T_c)$ & Average time to reach a checkpoint \\		
				%\hline
				%$G(T_c) \text{ or } Overhead(T_c)$ & $\frac{E(T_c)}{T_c}$ \\
				%\hline
				%$T$ & Total time to finish the job \\
				\hline
				$\lambda$ & Failure rate \\
				\hline
			\end{tabular}
			
			\label{tab:notation}
		\end{table}
	\fi
	
	To simplify our analysis, we make the following  assumptions:
	\begin{enumerate}
	%\textbf{Assumption: } For the sake of simplicity, we have taken the following three assumptions: 
	\item A faulty process can be restarted only after every checkpoint interval. This is a valid assumption in this context because Volpex server cannot detect a dead process immediately. Recruited client sends periodic heartbeat signals to the server. If server misses a particular client's heartbeat for a threshold amount of time, it assumes that the client is dead and reassigns the job to a new client. 
	\item Time to detect fault and restart is negligible. This is also a valid assumption in the context of Volpex because time to detect and restart is very small compared to the checkpoint interval. Rarely, a client failure occurs at the very last moment of the checkpoint interval period. If a client failure occurs even a few seconds earlier than the checkpoint interval, then the failure detection and restart time is overlapped by the remaining interval time. As a result, no extra time is wasted for fault detection and restart. Volpex does not have a fail-stop model like MPI where one process failure triggers other processes' immediate termination. Moreover, Volpex maintains a hot standby pool of spare nodes all the time so that a restart can be done as soon as a process is dead. 	
	\item All the nodes in volunteer network follow the same success probability distribution. While this is certainly not the case in practice, it is unrealistic to model each node individually in a heterogeneous volunteer environment. It is also unrealistic to have different checkpoint interval for different nodes. Our hypothesis is that an analysis based on a success distribution derived from the average MTTF for a pool of nodes will provide a realistic estimate of the optimal checkpoint interval.
	(Our experimental results validate this hypothesis.)
	\item  The success probability distribution is an {\em exponential distribution} function. We consider this as a valid assumption because the exponential distribution is commonly used to model failures in similar settings (e.g., \cite{Ross, Young:1974_FirstOrder}). 
	\end{enumerate}
	
	\subsection{Analysis for single process with single replica} 
	
\begin{figure} [h]
		\includegraphics[scale=0.33]{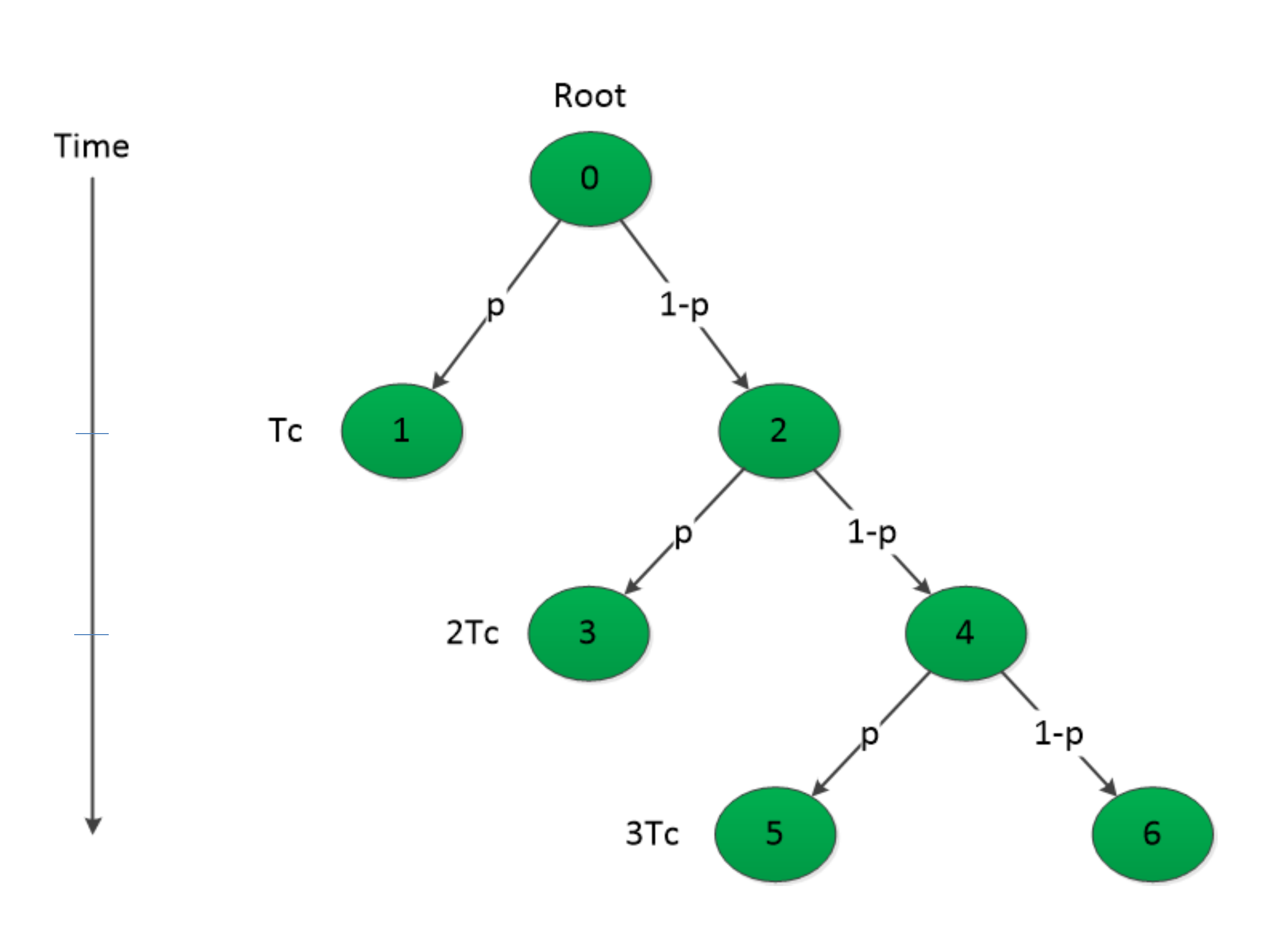}
		\caption{Illustration to calculate the expected overhead $G(T_c)$. Edges of the tree denote
			the probability to reach the next node and labels outside the nodes denote the time to reach that node from root.}
		\label{fig:tree}
	\end{figure}

	Assuming that the checkpoint interval is $T_c$, we calculate $G(T_c)$, the expected overhead. See  Figure~\ref{fig:tree} which uses a tree to explain the calculation.
	
	We  start from the root (node 0). After time $T_c$, the probability to reach the next checkpoint (node 1) successfully is given by $p = f(T_c) = e^{-\lambda T_c}$.  Since we assumed that a faulty process can be restarted only after every checkpoint interval period, if the process fails, after time $T_c$ we will end up in node 2. In that case we shall try to reach next checkpoint (node 3) in the next $T_c$ time. There is again the  probability $p$ to reach node 3 successfully after time $2T_c$. If the process fails again, we will be on node 4 at time $2T_c$. From node 4, we shall try to reach the next checkpoint (node 5) by $3T_c$ and so on. The edges of the tree denote the probability to reach the next node and labels outside the nodes denote the time to reach that node from root. For simplicity, we have the time labels outside the left nodes only. Note that once we successfully reach the end of a checkpoint interval,
	we spend $T_s$ time to create a checkpoint which contributes to the overhead.
	
	From the above discussion, for a single process with no replica, the expected overhead $G(T_c)$ is
	
	\iffalse
	
	\begin{multline}
		G(T_c) = p T_c + (1-p)p * 2T_c +(1 - p)2p * 3T_c \\+ (1-p)3p * 4T_c + ... 
		=  \frac{T_c}{p} \label{eq:3}
	\end{multline}

	Now consider we have 2 inter-communicating processes running parallel. If one process fails, the remaining process halts.
	For 2 processes (with no replica), probability to reach a checkpoint without failure $ = f(t)* f(t) = f(t)^2$
	
	So, For $n$ processes (with no replica), probability to reach a checkpoint without failure = $f(t)^n$
	
	For $n$ processes (with no replica), Average time to reach a checkpoint, 
	\begin{equation}
		E(T_c)= \frac{T_c}{f(T_c)^n}
	\end{equation}
	\\For $n$ processes (with no replica),
	\begin{equation}
		overhead(T_c) = \frac{E(T_c)}{T_c} = \frac{1}{ f(T_c)^n}
	\end{equation}
	
	As $T_s$ is not negligible, a process will require $T_s$ time to create the checkpoint after it successfully reach at a checkpoint; For a faulty process, it does not require $T_s$ time to create a checkpoint. Since $T_r$ is much smaller than $T_c$, By the time a process reaches to the next checkpoint interval period $T_c$, it can be detected as faulty by this time and will be ready for restart. 
	\fi
	%We consider $T$ is the total time to finish the job. If the optimal checkpoint interval is $T_c$ then it will take $\frac{T}{T_c}$ rounds to finish the job. %
	%Then for a single process with no replica, Average time to reach a checkpoint, 
	\begin{align}
	G(T_c)&=p(T_c+T_s)+(1-p)p(2T_c + T_s)+\notag\\
	&\phantom{{}=1}(1-p)^2p(3T_c+T_s) +(1-p)^3p(4T_c+T_s)+... \notag\\
 	&=\frac{T_c}{p} + T_s\notag\\
 	&=\frac{T_c}{f(T_c)} + T_s \text{ (replacing $p$ with $f(T_c)$)} \label{eq:6}
	\end{align}
	Thus the normalized overhead is:
	\begin{equation}
		\frac{G(T_c)}{T_c}  =\frac{1}{f(T_c)} + \frac{T_s}{T_c} \label{eq:7}
	\end{equation}

	To minimize the normalized overhead for a single process with no replica, we will differentiate  equation \ref{eq:7} with respect to $T_c$ and set the value to zero.	
	\begin{align*}
		&\frac{d}{dT_c} \frac{G(T_c)}{T_c}  =\frac{d}{dT_c}\frac{1}{f(T_c)} + \frac{d}{dT_c}\frac{T_s}{T_c} \notag\\
		&\text{or, } 0 = \frac{f'(T_c)}{f(T_c)^2} + \frac{T_s}{T_c^2}  
	\end{align*} 
	where $f(T_c) = e^{-\lambda T_c}$ (since we assume the exponential distribution) and $T_s$ is a constant.		
	\begin{align}
		%&\text{or, } 0 = \frac{f'(T_c)}{f(T_c)^2} + \frac{T_s}{T_c^2} \\
		&\text{or, } 0 = \frac{-\lambda e^{-\lambda T_c}}{({e^{-\lambda T_c})}^2} + \frac{T_s}{T_c^2} \notag\\
		%&\text{or, } 0 = \frac{-\lambda e^{-\lambda T_c}}{{e^{-2\lambda T_c}}} + \frac{T_s}{T_c^2} \notag\\
		%&\text{or, } 0 = -\lambda e^{\lambda T_c} + \frac{T_s}{T_c^2} \\
		%&\text{or, } \lambda e^{\lambda T_c} = \frac{T_s}{T_c^2} \\
		%&\text{or, } \ln (\lambda e^{\lambda T_c}) = \ln \frac{T_s}{T_c^2} \\ 
		%&\text{or, } \ln \lambda + \lambda T_c \ln e = \ln {T_s} - 2 \ln {T_c}
		&\text{or, } \lambda T_c + 2 \ln {T_c} = \ln {T_s} - \ln \lambda  \label{eq:10}
	\end{align}

	Using WolframAlpha's ~\cite{WolframAlpha} computational engines, we can solve equation \ref{eq:10} and get the optimal checkpoint interval $T^{opt}_c$ as follows:
	\begin{align}
		T^{opt}_c =  \frac{2W(  \frac{\sqrt{\lambda T_s}}{2} )}{\lambda} \label{eq:11} 
	\end{align}
	\iffalse
	\begin{align}
		T^{opt}_c =  \frac{2W( - \frac{\sqrt{\lambda T_s}}{2} )}{\lambda} \label{eq:12}  
	\end{align}
	\fi
	
	where $W(z)$ is the Lambert $W$ function. (The Lambert $W$ function  is the {\em inverse function} of the function $f(W) = W e^W$.)
	
	Using a first order approximation, $e^x  \approx 1 + x$, we can get a closed form formula for $T_c^{opt} = \sqrt{\frac{T_s}{\lambda}}$. Of course, using higher order approximations for the exponential function gives more accurate values
	of $T_c^{opt}$. Another option is to use compute $W$ function using mathematical software such as Mathematica.
	For example, if checkpoint size = 50KB, $\lambda$ = 0.0000348074 and $T_s$ = 1 sec then we get, $T_c^{opt}$ = 169 sec.

	\subsection{Generalization to multiple processes with many replicas}
	Next we will consider replication for each process. Suppose there are 2 replicas for each process. The process will restart only if both the replicas are dead.
	For a single process and without replica, probability to reach a checkpoint without failure = $p$.
	Then for a single process, probability to restart without finishing the job = $1-p$.
	If there are 2 replicas, probability to restart without finishing the job = $(1- p)*(1-p)  = (1-p)^2$.
	Hence, for a single process and with 2 replicas, the probability to reach a checkpoint without failure = $1- (1-p)^2$. Hence, we have that for a single process and with $r$ replicas, the probability to reach a checkpoint without failure = $1- (1-p)^r$. Generalizing, we have   for $n$  inter-dependent parallel processes and with $r$ replicas for each process, the probability for all the $n$ processes to successfully reach a checkpoint without failure = $(1- (1-p)^r)^n$. This is because, each of the $n$ processes will successfully reach a checkpoint, if at least one of the replicas of each process reaches a checkpoint.
	
	Thus, replacing $f(T_c)^n$ by $(1- (1-p)^r)^n$ in eq: \ref{eq:6}, we get, for $n$ processes each with $r$ replicas, the expected overhead is:
	\begin{equation}
		G(T_c) =\frac{T_c}{(1- (1-p)^r)^n} + T_s \label{eq:13}
	\end{equation}
	
	And normalized expected overhead is:
	\begin{equation}
		\frac{G(T_c)}{T_c} = \frac{1}{(1- (1-p)^r)^n} + \frac{T_s}{T_c} \label{eq:14}
	\end{equation} 
	
	To find the optimal checkpoint interval, we will replace $p$ by $f(T_c)$ in the above equation, differentiate it with respect to $T_c$ and then set the value to zero. 
	\begin{equation}
		0 = -n {(1- (1-f(T_c))^r)}^{-n-1} * \frac{d}{dT_c}(1-(1-f(T_c))^r) - \frac{T_s}{T_c^2}  \label{eq:15}
	\end{equation}
	
	\iffalse
	Again Let, probability distribution function of success distribution is an exponential distribution function. Then we can use Cumulative Distribution Function (CDF) to find the probability that a process is alive till time $t$.That is, 
	\begin{equation*}
		f(T_c) = e^{-\lambda T_c} \text{ where $\lambda$ is the failure rate}
	\end{equation*}
	\fi
	
	Setting $f(T_c) = e^{-\lambda T_c}$  in eq \ref{eq:15} we get, 
	\begin{align}
		 0 &= -n {(1- (1-e^{-\lambda T_c})^r)}^{-n-1} \notag\\ &\phantom{{}=1}*\frac{d}{dT_c}(1-(1-e^{-\lambda T_c})^r) - \frac{T_s}{T_c^2}  \notag\\
		\text{or, }0 &= n {(1- (1-e^{-\lambda T_c})^r)}^{-n-1} \notag\\ &\phantom{{}=1} * r\lambda e^{-\lambda T_c} (1-e^{-\lambda T_c})^{r-1} - \frac{T_s}{T_c^2} \notag\\
		\text{or, } \frac{T_s}{T_c^2} & = \frac{n r\lambda e^{-\lambda T_c} (1-e^{-\lambda T_c})^{r-1}}{{(1- (1-e^{-\lambda T_c})^r)}^{n+1}} \label{Final equation for CPI}
	\end{align}
	
	\iffalse
	Now, 
	\begin{align*}
		\frac{d}{dT_c}(1-(1-e^{-\lambda T_c})^r) &=-r\lambda e^{-\lambda T_c} (1-e^{-\lambda T_c})^{r-1}
		%\frac{d}{dT_c}(1-(1-e^{-\lambda T_c})^r) &= -\frac{d}{dT_c}(1-e^{-\lambda T_c})^r\\
		%=-r (1-e^{-\lambda T_c})^{r-1} (-\frac{d}{dT_c}(e^{-\lambda T_c}))\\
		%&=-r (1-e^{-\lambda T_c})^{r-1} ((e^{-\lambda T_c}) \lambda)\\
		%&=-r\lambda e^{-\lambda T_c} (1-e^{-\lambda T_c})^{r-1}
	\end{align*}
	
	Replacing the above value in equation \ref{eq:16} we get, 
		
	\begin{align}
		%& 0 = -n {(1- (1-e^{-\lambda T_c})^r)}^{-n-1}  *(-r\lambda e^{-\lambda T_c} (1-e^{-\lambda T_c})^{r-1}) - \frac{T_s}{T_c^2} \notag\\
		&\text{or, }0 = n {(1- (1-e^{-\lambda T_c})^r)}^{-n-1} \notag\\ &\phantom{{}=1} * r\lambda e^{-\lambda T_c} (1-e^{-\lambda T_c})^{r-1} - \frac{T_s}{T_c^2} \notag\\
		&\text{or, } \frac{T_s}{T_c^2} = \frac{n r\lambda e^{-\lambda T_c} (1-e^{-\lambda T_c})^{r-1}}{{(1- (1-e^{-\lambda T_c})^r)}^{n+1}} \label{Final equation for CPI}
	\end{align}
	\fi
	
	 Solving eq \ref{Final equation for CPI}, we can get optimal checkpoint interval $T^{opt}_c$.
	As an example,  if $r = 1$ from equation \ref{Final equation for CPI} we get, 
	\begin{align}
		%\frac{T_s}{T_c^2} &= \frac{n \lambda e^{-\lambda T_c}}{{(1- (1-e^{-\lambda T_c}))}^{n+1}} \notag\\
		%\text{or, }\frac{T_s}{T_c^2} &= \frac{n \lambda e^{-\lambda T_c}}{{(e^{-\lambda T_c})}^{n+1}} \notag\\
		\frac{T_s}{T_c^2} &= n \lambda e^{n \lambda T_c} \label{eq:18}
	\end{align}
	
	By solving equation \ref{eq:18} we can get the optimal checkpoint interval $T_c$ for $r=1$ as below:
	\begin{align}
		T_c =  \frac{2W( \frac{1}{2} \lambda n \sqrt{ \frac{1}{\lambda n T_s} } T_s) } {\lambda n} \label{final eq for r=1.1}
	\end{align}
	
	\iffalse
	\begin{align}
		T_c =  \frac{2W(- \frac{1}{2} \lambda n \sqrt{ \frac{1}{\lambda n T_s} } T_s) } {\lambda n} \label{final eq for r=1.2} 
	\end{align}
	\fi
	
	Where $W(z)$ is the Lambert $W$ function.
	As a numerical example, if checkpoint size = 50 KB, $\lambda$ =  0.0000348074 and $T_s$ = 1 seconds then for $n=16$ we get, $T^{opt}_c \approx$ 42 seconds and for $n=32$ we get, $T^{opt}_c \approx$ 29 seconds.
	For $r>1$, we can get $T^{opt}_c$ by solving equation \ref{Final equation for CPI} using WolframAlpha's computational engines.

	\section{Experiments and Results}
The theoretical results developed in this paper predict the checkpoint interval that minimizes the execution time, but they are based on several assumptions that are, at best, approximations of real world behavior. The experimental evaluation in this paper is to determine if the predicted optimal checkpoints are practically useful, despite the imperfect assumptions.

	\subsection{Platform for experiments}
	Experiments were performed on a testbed of volunteer nodes managed by BOINC and Volpex.
	The testbed consisted of approximately 350 active volunteer nodes. Approximately
	20\% of the nodes were on the university campus and the remaining 80\% were distributed worldwide. For the purpose of evaluation, Replica Exchange Molecular Dynamics (REMD)~\cite{DirarHomouz08192008,LorenStagg11272007,qian11,samiotakis10} code was executed repeatedly. This application consists of a set of independent processes that coordinate and communicate regularly using the Volpex Dataspace API. The application is instrumented to include flexible checkpointing; every application process makes frequent checkpoint requests, but an actual checkpoint is stored only when a request is accepted by the server. Hence the server can be configured to define the actual checkpoint interval.
	
	\subsection{Measurement of parameters for checkpoint prediction}
	
	%The algorithm is implemented in C++ inside Volpex server source code. 
	The key system parameters for estimating the optimal checkpoint interval are the failure distribution that is based on mean time to failure (MTTF) and the time to record a checkpoint. We recorded all failures of our system nodes for a substantial duration of time. Note that failure of a volunteer node can be caused simply by a system being turned off by the owner or a temporary loss of network activity, not just software or hardware faults. We estimated the MTTF for the nodes  in our volunteer pool using the following equation:
	
	\begin{equation}
		MTTF = \frac{\text{Total hours of operation}}{\text{Total Number of failures}} \label{MTTF equation}
	\end{equation}

	 For the measurement made on our system, we get:
	
	\begin{align*}
	\mbox{Total Hours of Operation } & = 13678.67\\
	\mbox{Total Number of Failures } & = 1714\\
	\mbox{MTTF }= 7.9805 hrs &= 28730 sec\\
	\lambda  = \frac{1}{MTTF} & = 0.0000348074
	\end{align*}

	From Equation~\ref{Ts formula}, we know that the checkpointing time $T_s$ includes the time to save the checkpoint on the client, the network delay to send the checkpoint information from user PC to the checkpoint server, the time to upload checkpoint on the server and time to send acknowledgment to the client. Experiments were conducted with 50 KB and 5 MB checkpoint sizes. %In general for 50 KB checkpoints, $T_{s(cl)}$ is negligible compared to other 3 components. 
	%Saving such a small size checkpoint in the server is almost negligible compared to the network latency. 
	The network delay depends on a number of factors including the network latency between the client PC and the server, network congestion, and link bandwidth. The time taken for the creation of a checkpoint on the server can also vary depending on the workload on the server. In our experiments, we recorded the time to save every checkpoint in the server in a log file. We observed that $T_s$ varies between 200 ms to 1200 ms for 50 KB checkpoints and between 1 second to 290 seconds for 5 MB checkpoints. However, when multiple processes create independent checkpoints, the maximum checkpoint interval is the time for which the overall computation is stalled. 
	
	To find the $T_s$ value from the recorded log, we computed the average of the maximum of $n*r$ random readings from the log file ($n$ = number of process and $r$ = number of replica). This computation was performed by iteratively selecting a large set of random readings and computing the average.  
	
	For	50 KB checkpoints, we employed $T_s$ = 1 second as the estimated time to create a checkpoint. For 5 MB checkpoints, we estimated $T_s$ = 156, 187 and 212 s for 16, 32 and 64 processes respectively. Please note that we used one $T_s$  value for different $n*r$ values for 50 KB checkpoints since they differ very little. Finally, the predicted optimal checkpoint intervals were computed from the model developed in section \ref{sec:model} using WolframAlpha's~\cite{WolframAlpha} computational engines.
	
	\subsection{Experimental results }	
	
	\begin{figure*}
		\includegraphics[width=\textwidth]{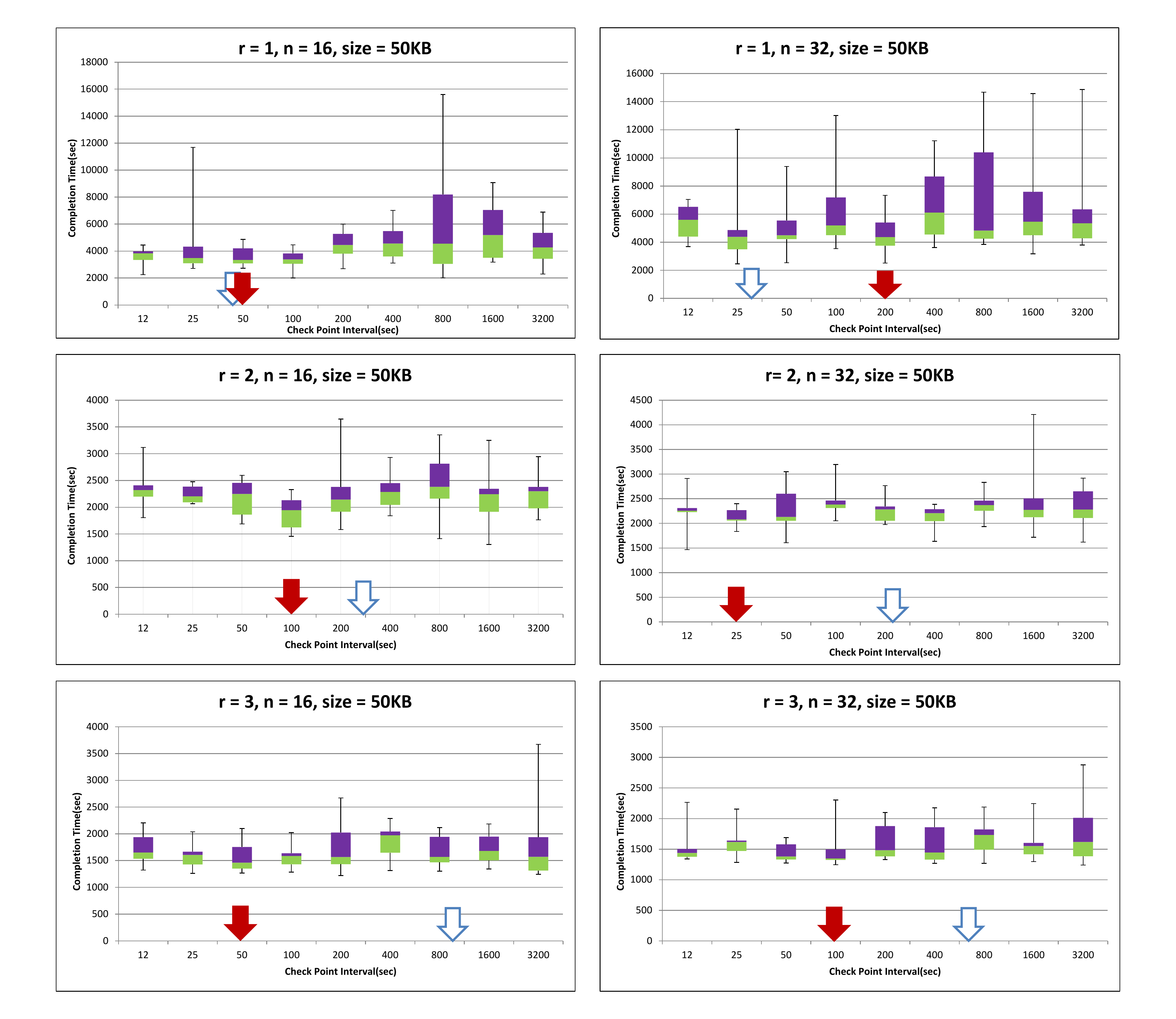}
		\caption{Completion times for varying checkpoint intervals in different execution scenarios with 50 KB checkpoints. The box plot shows the lowest, 25th quartile(bottom of green box), median, 75th quartile(top of purple box), and the highest execution times over multiple runs. Solid red arrow indicates the checkpoint interval with the lowest median completion time, while the blue blank arrow indicates our predicted optimal checkpoint interval.}
		\label{fig:boxplot_50kb}
	\end{figure*}
	
	\begin{figure*}
		\includegraphics[width=\textwidth]{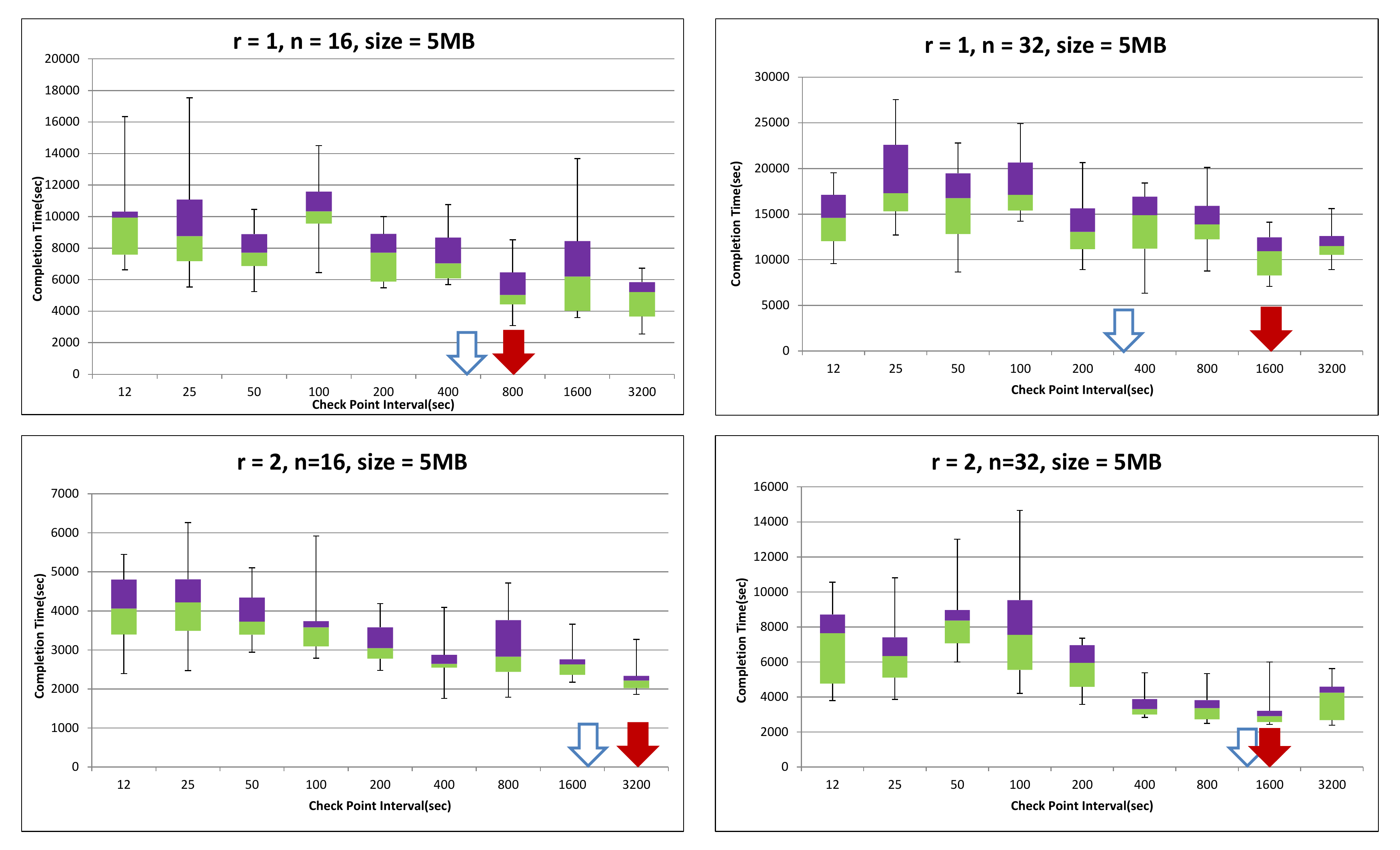}
		\caption{Completion times for varying checkpoint intervals in different execution scenarios with 5 MB checkpoints. The box plot shows the lowest, 25th quartile(bottom of green box), median, 75th quartile(top of purple box), and the highest execution times over multiple runs. Solid red arrow indicates the checkpoint interval with the lowest median completion time, while the blue blank arrow indicates our predicted optimal checkpoint interval.}
		\label{fig:boxplot_5mb}
	\end{figure*}
		
	We executed the REMD code on 16 and 32 processes with no replicas, 2 replicas for each process and 3 replicas for each process. When going from 16 to 32 nodes, the problem size is scaled up (doubled) such that the computation workload per process is unchanged. The checkpoint sizes were 50 KB and 5 MB. A fixed checkpoint interval between 12 to 3200 seconds was employed for each run and the completion times were recorded. Each experiment was run ten times.  Note that it is very important to have multiple runs for volunteer nodes as the execution time is expected to vary significantly between the runs based on the failure patterns of nodes in a run as well as the CPU and network characteristics of the specific sets of nodes selected for execution in each run.      
	
	Once we completed all the experiments, we picked the lowest median completion time from each scenario and the corresponding heuristically computed optimal checkpoint interval. We used these lowest completion times and the corresponding optimal checkpoint intervals for performance evaluation.

	The pattern of completion time for varying checkpoint intervals is illustrated in Figures~\ref{fig:boxplot_50kb} and ~\ref{fig:boxplot_5mb} with different levels of replication. These box plots represent the results as quartiles of execution times, including the minimum, maximum and the median execution time. Substantial variation is observed between execution runs due to the nature of a volunteer computing platform. We observe that in Figure~\ref{fig:boxplot_50kb}, minimum completion time is achieved with a lower checkpoint interval compared to Figure~\ref{fig:boxplot_5mb}. The reason is that, with a larger checkpoint size, it is often better to save a checkpoint less frequently. We note that there is often a significant difference between the predicted optimal checkpoint interval and the checkpoint interval for which the lowest median execution time is recorded. However, we note that the actual execution time with the predicted optimal checkpoint interval is typically very close to the measured lowest median execution time. This is examined further in this section.

	\begin{table} [ht]
		\centering
		\resizebox{\columnwidth}{!}{
			\begin{tabular}{|c|c|c|c|c|c|c|c|c|}
				\hline
				\multicolumn{1}{|p{1.1cm}|}{\centering Checkpoint \\size} & $n$ &  $r$ &
				\multicolumn{1}{|p{1.1cm}|}{\centering Median \\of lowest \\ completion \\ time,\\ $T_{best}$ \\ (s)} & 
				\multicolumn{1}{|p{1.1cm}|}{\centering Median \\of highest\\completion \\ time,\\$T_{worst}$ (s)} & 
				\multicolumn{1}{|p{1.1cm}|}{\centering Our \\predicted \\Checkpoint\\Interval, \\ $T_c$ \\(s)}  & 
				\multicolumn{1}{|p{1.1cm}|}{\centering Calculated \\ completion\\ time\\for $T_c$,  \\ $T_{predict}$\\(s)}  &
				\multicolumn{1}{|p{1cm}|}{\centering Abs. \% of \\difference \\between \\$T_{best}$ and \\$T_{predict}$  \\ (\%)} & 
				\multicolumn{1}{|p{1cm}|}{\centering Abs. \% of \\difference \\between \\$T_{best}$ and \\$T_{worst}$  \\ (\%)}  \\
				\hline
				50 KB & 16 & 1 &  3344 &  5191 & 42  & 3388  & 1.34  &  55.26  \\
				50 KB & 16 & 2 &  1946 &  2383 & 297 & 2213  & 13.74 &  22.43   \\
				50 KB & 16 & 3 &  1463 &  1973 & 851 & 1576  & 7.74  &  34.87  \\
				50 KB & 32 & 1 &  4362 &  6108 & 29  & 4398  & 0.82  &  40.03  \\
				50 KB & 32 & 2 &  2081 &  2388 & 235 & 2270  & 9.11  &  14.52   \\		
				50 KB & 32 & 3 &  1348 &  1732 & 714 & 1671  & 23.93 &  28.49   \\
				 5 MB & 16 & 1 &  5028 &  10333& 465 & 6710  & 33.45 &  105.50   \\
			 5 MB & 16 & 2 &  2217 &  4222 & 1708& 2604  & 17.50 &  90.46   \\
			 5 MB & 32 & 1 &  10934&  17298& 339 & 14331 & 31.08 &  58.21   \\
			 5 MB & 32 & 2 &  2906 &  18374& 1398& 3020  & 3.94  &  188.19  \\		
				\hline
			  &    &   &       &       &     & Average& 14.26&  63.79 \\
				\hline
			  &    &   &       &       &     & Max    & 33.45& 188.19\\	
				\hline
			\end{tabular}
		}
		\caption{Performance comparison among measured best,  worst and our predicted optimal checkpoint interval}
		\label{tab:Performance comparison}
	\end{table}

	Table \ref{tab:Performance comparison} presents the application performance with predicted checkpoint intervals, along with the best and worst measured performance over the entire range of checkpoint intervals. Linear interpolation was employed  for finding the job completion time for the predicted optimal checkpoint interval when the interval was between points for which measurements were made. 
	The percentage difference between the lowest completion time and completion time with the predicted optimal checkpoint interval is the performance prediction error.
	
	We observe from Table \ref{tab:Performance comparison} that percentage difference between the lowest measured completion time and completion time with the predicted optimal checkpoint interval (prediction error) ranges between 0.82\% and 33.45\% with an average prediction error of 14.26\%. In a volunteer framework this is a low error given the inherent performance variations in a volunteer framework. Hence, the predicted checkpoints are practically useful; in other words only an improvement of around 14\% is expected with an exhaustive search for the best checkpoint interval.
	
	Table \ref{tab:Performance comparison} also lists the percentage difference between the highest(worst) measured completion time and the completion time with the predicted optimal checkpoint interval, which  ranges between 14.52\% and 188.19\% with an average difference of 63.79\%. The point is that selecting a good checkpoint interval is important and selecting a poor checkpoint interval heuristically can have substantial implications for performance. The predicted checkpoint interval always leads to a performance fairly close to the optimal performance, and far superior than the performance with a poorly selected checkpoint interval.

	%Please note that current Volpex framework uses a heuristic predefined checkpoint interval. As a result we run the risk of ending up with a checkpoint interval that can cause the worst completion time, $T_{worst}$. We compared the   performance of our approach with that of $T_{worst}$ and found that on an average, ours is better by 44.76\%  with a maximum of 177.28\% performance increase. We also observe that the performance difference between best and worst completion time can be a large as 188.19\% with a 63.79\% average, which can be minimized by selecting an optimal checkpoint interval.
	
	\begin{figure}
		\includegraphics[width=\columnwidth]{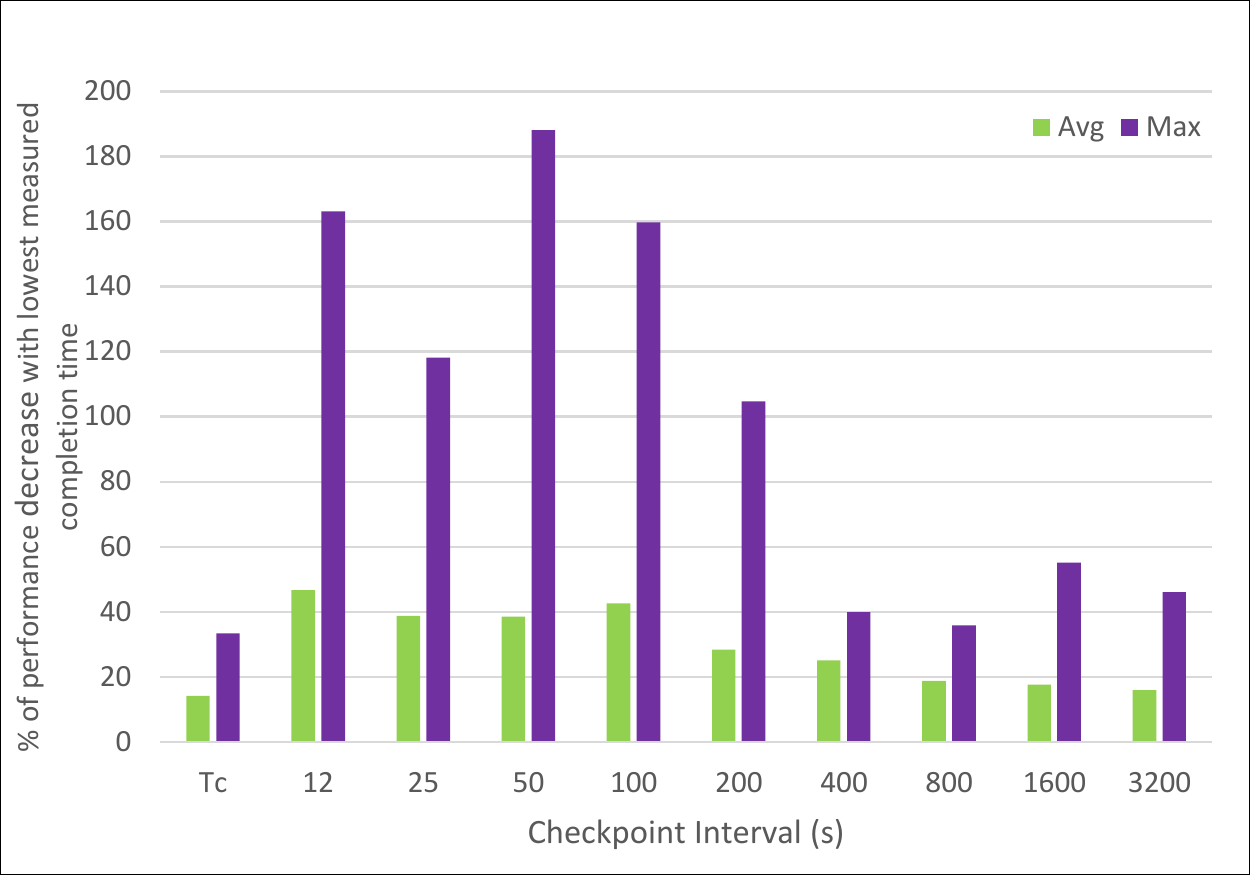}
		\caption{Performance penalty with always using a predefined fixed checkpoint interval versus the performance penalty using our predicted optimal checkpoint interval ($T_c$).}
		\label{fig:performance fixed bar}
	\end{figure}
	
	Figure~\ref{fig:performance fixed bar} shows the performance penalty with employing a fixed predefined checkpoint interval for all scenarios in comparison with the measured lowest completion time in each scenario. It also shows the performance penalty with a predicted optimal checkpoint interval (first set of bars) for all scenarios in comparison with the measured lowest completion time in each scenario. Clearly using the predicted optimal checkpoint interval fares much better on average than employing any predefined checkpoint interval all the time. Moreover, even for the worst case, employing a predicted checkpoint yields a performance within ~33\% of the best measured performance, whereas it ranges between ~39\% and ~188\% for a fixed checkpoint interval.
	
	Hence our predicted checkpoint interval is not truly optimal but it leads to an execution time close to the minimum possible job execution time. We did not know the true optimal checkpoint interval that results in the lowest completion time beforehand as it is determined empirically. The goal was to predict a checkpoint interval that yields an execution time close to the optimal. Given that the performance difference with two different checkpoints can be 188\% or higher, we  claim that the predicted checkpoint interval is an excellent choice in practice. We do expect some error in identifying the optimal checkpoint interval given the nature of the volunteer environment where the execution performance typically varies significantly between runs. Also, this work makes a number of pragmatic assumptions detailed earlier that are the likely reasons for suboptimal experimental results.

	%The result shows that job executed with Optimal CPI needs smaller completion time and thus performs better than any arbitrary CPI. If we consider the median value, we get significant performance improvement compared to fixed arbitrary CPI. In some cases the improvement between best and worst case is more that 200\%. We observed that almost every time, with our calculated CPI, the job completion time is close to minimum job completion time.  

	%\input{futurework.tex}
	
	\balance
	
	\section{Concluding Remarks}
	
	This paper introduces a  mathematically-based methodology to estimate a checkpoint interval that minimizes completion time for communicating inter-dependent parallel processes running in a high failure volunteer environment employing checkpointing and replication for fault tolerance. The results of the model are evaluated with the Volpex execution framework by running a real world application with various checkpoint size and replication. The results indicate that the model is effective in identifying a checkpoint interval that leads close to best possible application performance in this environment. %The results indicate that the model is effective in identifying a checkpoint interval that leads to application performance that is close to the best that is possible in the environment. 
	Since the true optimal checkpoint interval can only be determined empirically after extensive experimentation, a predicted checkpoint interval is the only reasonable choice, which is much better than taking an arbitrary value.
	
	%We believe that this is the first research project that addresses the optimization of checkpointing for communicating parallel processes that must employ checkpointing and process replication for effective execution because of the high node failure rates. 
	We believe, this is the first research project which addresses the optimization of checkpointing for communicating parallel processes that must employ checkpointing and process replication for effective execution because of the high node failure rates. A number of assumptions were made in this work regarding the failure distribution of nodes and the modeling of execution with checkpointing and replication. Future work will generalize the checkpoint interval prediction model. However, we believe, the current model can predict checkpoint intervals that are valuable in practice. Future work will also address evaluation in a larger number of scenarios and different size checkpoints.
	
	Finally, while the checkpoint prediction model is evaluated in a volunteer environment, this work presents a fundamental contribution towards fault tolerant distributed systems. The results are applicable to other distributed systems such as computation clouds where replication and checkpointing are used together to gain acceptable performance in the face of failures.
	
	%The algorithm minimizes overall completion time by adopting the optimal checkpoint interval. It trades off between resource overhead due to checkpoints creation and  work loss due to client failure. The algorithm is designed based on the statistical property of the available clients. We implemented our algorithm in Volpex execution framework and tested it using a real world application. The algorithm minimizes overall completion time compared to previous fixed checkpoint interval system and shows a significant improvement to increase the throughput of the system. Although our model is designed and tested in a volunteer environment, the concept can also be extended to a distributed network. 	

	\bibliographystyle{IEEEtranS}
	\bibliography{reference}	
	
	\section*{}
	\pagebreak

	%\nobalance 
	
\end{document}